\documentclass{aa}
\usepackage{graphicx}
\usepackage{txfonts}
\usepackage{natbib}
\usepackage{color}
\usepackage[colorlinks, allcolors=blue]{hyperref}
\newcommand{\HRIEUV}{HRI$_{\mathrm{EUV}}$}

\begin{document}

\title{A multi-instrument study of ultraviolet bursts\\ and associated surges in AR 12957}

\author{C. J. Nelson$^{1}$, D. Calchetti$^{2}$, A. Gandorfer$^{2}$, J. Hirzberger$^{2}$, J. Sinjan$^{2}$,\\ S. K. Solanki$^{2}$, D. Berghmans$^{3}$, H. Strecker $^{4}$, J. Blanco$^{5}$}

\offprints{chris.nelson@esa.int}
\institute{$^1$European Space Agency (ESA), European Space Research and Technology Centre (ESTEC), Keplerlaan 1, 2201 AZ, Noordwijk, The Netherlands\\
$^2$Max Planck Institute for Solar System Research, Justus-von-Liebig-Weg 3, 37077 G\"ottingen, Germany\\
$^3$Solar-Terrestrial Centre of Excellence – SIDC, Royal Observatory of Belgium, Ringlaan -3- Av. Circulaire, 1180 Brussels, Belgium\\
$^4$Instituto de Astrofísica de Andalucá (IAA-CSIC), Apartado de Correos 3004, E-18080 Granada, Spain\\
$^5$Universitat de València, Catedrático José Beltrán 2, 46980 Paterna-Valencia, Spain}

\date{}

\abstract
{The relationship between ultraviolet (UV) Bursts and solar surges is complex, with these events sometimes being observed together and sometimes being observed independently. Why this sporadic association exists is unknown, however, it likely relates to the physical conditions at the site of the energy release that drives these events.}
{Here, we aim to better understand the relationship between UV Bursts and solar surges through a multi-instrument analysis of several associated events that occurred around the trailing sunspot in AR $12957$.}
{We use data from Solar Orbiter, the Solar Dynamics Observatory (SDO), and the Interface Region Imaging Spectrograph (IRIS) to achieve our aims. These data were sampled on 3rd March 2022 between $09$:$30$:$30$ UT and $11$:$00$:$00$ UT, during which time a coordinated observing campaign associated with the Slow Solar Wind Connection Solar Orbiter Observing Plan (SOOP) took place.}
{Numerous small-scale negative polarity magnetic magnetic features (MMFs) are observed to move quickly (potentially up to $3.3$ km s$^{-1}$) away from a sunspot until they collide with a more stable positive polarity plage region around $7$ Mm away. Several UV Bursts are identified in IRIS slit-jaw imager (SJI) $1400$ \AA\ data co-spatial to where these opposite polarity fields interact, with spatial scales ($2$ Mm$<$) and lifetimes ($20<$ minutes) larger than typical values for such events. Two surges are also observed to occur at these locations, with one being short ($5$ Mm) and hot (bright in IRIS SJI images), whilst the other is a cooler (dark in coronal imaging channels), longer surge that appears to fill an active region loop.}
{Magnetic reconnection between the negative polarity MMFs around the sunspot and the positive polarity plage region appears to be the driver of these events. Both the speed of the MMFs and the locally open magnetic topology of the plage region could possibly be important for forming the surges.}

\authorrunning{Nelson}
\titlerunning{Multi-Instrument Study Of UV Bursts And Associated Surges}

\maketitle

\section{Introduction}
\label{Introduction}

Small-scale bursts, typically with diameters less than $5$ Mm and lifetimes of less than $10$ minutes, appear to be ubiquitous within the solar atmosphere. Spectroscopic imaging of the transition region from, for example, the Interface Region Imaging Spectrograph (IRIS; \citealt{DePontieu14}) reveals features named ultraviolet (UV) Bursts (\citealt{Young18}), that occur seemingly across the entire solar disk with temperatures close to $10^5$ K. These events can be further split into sub-categories such as IRIS bursts (\citealt{Peter14}) and Explosive Events (EEs; \citealt{Brueckner83}) if spectral information in the UV is sampled co-spatial to them. The connectivity between UV Bursts and similar features in other regions of the solar atmosphere is currently an area of focused research in the field. At lower temperatures, UV Bursts have been reported to form co-spatial to both Ellerman bombs (EBs; \citealt{Ellerman17}) and Quiet-Sun Ellerman-like Brightenings (QSEBs; \citealt{Rouppe16}), that have temperatures below $10^4$ K (\citealt{Vissers15,Nelson16}). At higher temperatures, recent research has shown UV Bursts forming co-spatially with the extremely small, short-lived extreme UV (EUV) brightenings (\citealt{Berghmans21,Nelson23}) that were one of the key early discoveries of Solar Orbiter's (\citealt{Muller20}) Extreme Ultraviolet Imager (EUI; \citealt{Rochus20}). 

\begin{figure}
\includegraphics[width=0.49\textwidth]{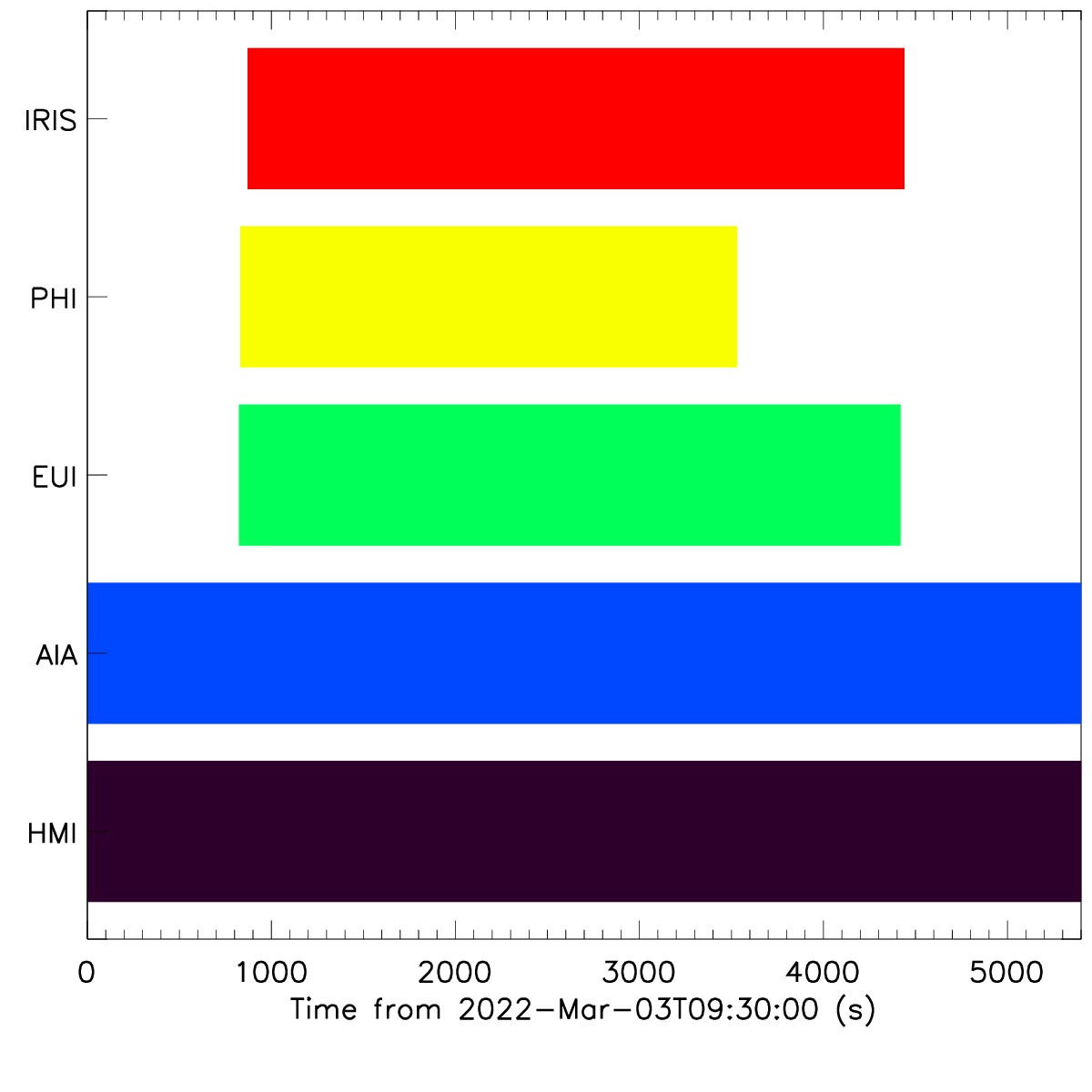}
\caption{Data availability for the five instruments analysed in this article. The filled bars indicate the times during which each instrument was sampling data, with the bars corresponding to EUI and PHI indicating, specifically, that the \HRIEUV\ and SO/PHI-HRT telescopes were operating. A more detailed summary of these data can be found within Sect.~\ref{DataProducts}}
\label{Data_avail}
\end{figure}

Numerous authors have hypothesised that magnetic reconnection could be the driver of small-scale bursts (see, for example, \citealt{Young18} and references therein), with the spectroscopic visibility of the events potentially being related to the local magnetic field topology and plasma beta (as discussed by \citealt{Peter19}). Some events (such as EBs and IRIS bursts) are typically modelled as being driven by magnetic reconnection occurring at multiple heights along extended current sheets situated low down in the solar atmosphere between regions of near-vertical field of opposite polarities (\citealt{Hansteen19}). Indeed, line-of-sight magnetic field maps provided by many instruments have revealed interactions between opposite polarity magnetic fields in the photosphere co-spatial to these events, with magnetic cancellation (the reduction in unsigned magnetic flux through time) often being measured  (e.g., \citealt{Peter14,Nelson16,Reid16} amongst many others). Other events (such as EEs and EUV brightenings) may, instead be driven by reconnection between nearly horizontal fields higher up in the solar atmosphere (\citealt{Innes99,Chen21}) and, as such, the co-spatial line-of-sight magnetic field in the photosphere exhibits little change during their evolution (e.g., \citealt{Nelson23}). The higher spatial resolution magnetic field maps provided by Solar Orbiter's Polarimetric and Helioseismic Imager (PHI; \citealt{Solanki20}) offer an excellent opportunity to investigate the evolution of the photospheric field co-spatial to localised brightenings in the solar atmosphere.

In addition to linkages with different manifestations of small-scale bursts, UV Bursts have also been associated with a range of other features in the solar atmosphere. One of the most common features observed to form together with UV Bursts are solar surges (see, for example, \citealt{Madjarska09,Nobrega17,Nobrega21}), with typical lengths up to $100$ Mm and lifetimes less than one hour. Surges, similarly to UV Bursts, are possibly driven by magnetic reconnection, specifically, interactions between emerging flux and the ambient field in the solar atmosphere (\citealt{Yokoyama95,Guglielmino10}), with either hot (transition region temperatures) or cool (chomospheric temperatures) plasma being driven upwards along the magnetic field lines. Why some UV Bursts form with surges, whilst others do not, is still an open question in the field. To make progress towards understanding this, data from multiple instruments would be required in order to build up an adequately complete picture of the evolution of both the local magnetic field and plasma (across multiple temperature windows).

In this article, we analyse a grouping of UV Bursts and associated surges that occurred close to the trailing positive polarity sunspot in AR 12957. We study the spectroscopic signatures of these events, from the photosphere to the corona, and investigate how these correlate to the evolution of the photospheric magnetic field. Our article is structured as follows: In Sect.~\ref{Observations} we introduce the data products studied here and the methods used to align data from different satellites; in Sect.~\ref{Results} we present the results obtained through our analysis; in Sect.~\ref{Discussion} we present a discussion of the implications of this analysis; before, finally, in Sect.~\ref{Conclusions} we draw our conclusions.

\section{Observations}
\label{Observations}

\subsection{Data products}
\label{DataProducts}

In this article, we analyse data sampled on $3$rd March $2022$ by three separate satellites, namely IRIS, Solar Orbiter, and the Solar Dynamics Observatory (SDO). Specifically, we focus on a 90-minute window between $09$:$30$:$00$ UT and $11$:$00$:$00$ UT within which the high-resolution remote sensing instruments on-board Solar Orbiter were conducting coordinated observations of AR $12957$ with IRIS for an instance of the `Slow Solar Wind' Solar Orbiter Observing Plan (SOOP). See \citet{Zouganelis20} for a summary of the scientific aims of this SOOP. During these coordinated observations, Solar Orbiter was close to the Sun-Earth line (-4$^\circ$ latitude, -6$^\circ$ longitude) and at a distance from the Sun of $0.547$ AU, leading to a time-$\Delta$ of $221$ s between light reaching Solar Orbiter and the Earth. For ease, all times reported for Solar Orbiter in the remainder of this article have been corrected for this time-$\Delta$ to the reference frame of Earth. A graphic summary of data availability for each instrument during the studied time window can be found in Fig.~\ref{Data_avail}, whilst a more detailed description of the experiments conducted by each instrument can be found below.

\begin{figure*}
\includegraphics[width=0.99\textwidth]{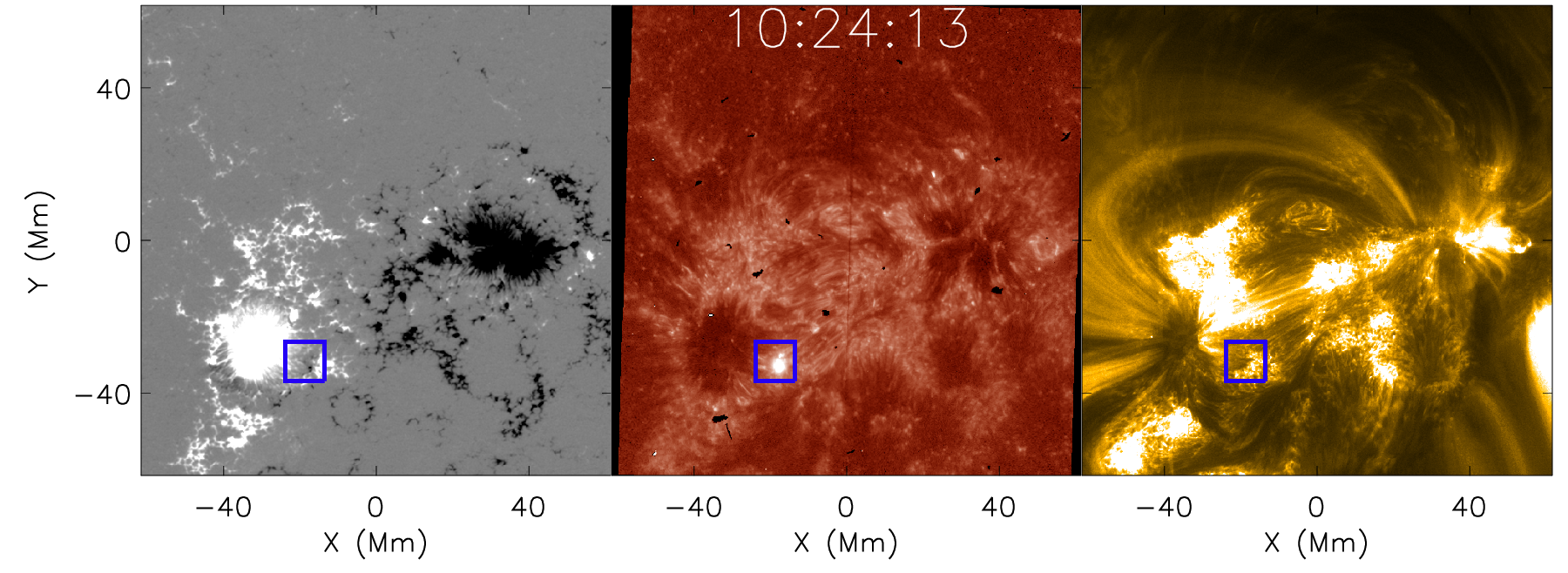}
\caption{Structuring of AR $12957$ in the line-of-sight magnetic field as observed by SO/PHI-HRT (left panel), the transition region as observed by the IRIS SJI $1400$ \AA\ filter (middle panel), and the corona as observed by the \HRIEUV\ telescope (right panel). Each panel plots the frame sampled by each instrument closest to $10$:$24$:$13$ UT. The SO/PHI-HRT line-of-sight magnetic field map is saturated at $\pm500$ G, with white pixels indicating positive polarity and black pixels indicating negative polarity. The over-laid blue box indicates the FOV surrounding  the grouping of UV Bursts, located within the flow region around the trailing positive polarity sunspot, plotted in Figs.~\ref{Mag_evol},\ref{LA_evol}-\ref{UA_evol}.}
\label{Overview}
\end{figure*}

\subsubsection{IRIS}
\label{IRIS}

Between $09$:$44$:$30$ UT and $10$:$44$:$09$ UT, IRIS ran a dense 4-step raster, with the spectrograph slit crossing the core of the active region (between the leading negative polarity sunspot and the trailing positive polarity sunspot) in the North-South direction. The raster step-size was $0.35$\arcsec\ ($254$ km), the length of the slit was $175$\arcsec\ ($127$ Mm), and the spectral sampling per pixel was $0.05$ \AA\ for all spectral windows. The exposure time of these spectral data was $4$ s, whilst the step cadence was $5.1$ s giving a total raster cadence of $20.4$ s. The slit-jaw imager (SJI) collected images sequentially in both the \ion{Si}{IV} $1400$ \AA\ and \ion{Mg}{II} $2796$ \AA\ channels, with cadences of $10.2$ s and pixel scales of $0.33$\arcsec\ ($239$ km). The total field-of-view (FOV) of the SJI data was around $122$$\times127$ Mm$^2$, centred at helioprojective-cartesian coordinates of $x_\mathrm{c}$=$-277$\arcsec, $y_\mathrm{c}$=$-129$\arcsec. The FOV as sampled by the $1400$ \AA\ SJI filter at $10$:$24$:$13$ UT is plotted in the middle panel of Fig.~\ref{Overview}, with the blue box highlighting the grouping of UV bursts in the flow region around the sunspot that are analysed in this article. The IRIS OBSID of this dataset is: 3600607418.

\subsubsection{Solar Orbiter}
\label{SolarOrbiter}

To complement these IRIS data, we also study data sampled by two instruments on-board Solar Orbiter. Firstly, we analyse maps of the coronal plasma structuring as sampled by EUI's High-Resolution Imager (\HRIEUV) telescope using its EUV $174$ \AA\ filter. We analyse L2 fits files, that are included in EUI data release 6.0\footnote{EUI data release 6.0: https://doi.org/10.24414/z818-4163}. The \HRIEUV\ telescope operated between $09$:$43$:$43$ UT and $10$:$43$:$38$ UT with pixel scales of $0.49$\arcsec\ at Solar Orbiter (which corresponds to a value of $196$ km on the Sun), exposure times of $2.8$ s, and cadences of $5$ s (returning a total of $720$ images). These data had a slight roll angle as compared to IRIS of approximately $-1.5^\circ$ which was apparent when considering the whole FOV, but not on the small-scales studied here. The $2048$$\times2048$ pixel camera of \HRIEUV\ covered a total FOV of around $401$$\times401$ Mm$^2$ and, as such, a smaller $701$$\times701$ pixel sub-field approximately matching the IRIS FOV was extracted for analysis here. Jitter within the \HRIEUV\ time-series was removed using the $tr\_get\_disp.pro$ IDL routine which worked qualitatively well. Any residual jitter (seemingly $\le1$ pixel) evident during the alignment of the different datasets was removed manually using the methods described in Sect.~\ref{Alignment}. An overview of the \HRIEUV\ FOV as observed at around $10$:$24$:$13$ UT can be found in the right-hand panel of Fig.~\ref{Overview}.

The second dataset provided by Solar Orbiter that is studied here is a time-series of line-of-sight magnetic field measurements returned by PHI's High-Resolution Telescope (SO/PHI-HRT; \citealt{Gandorfer18}). These data were sampled between $09$:$43$:$51$ UT and $10$:$28$:$51$ UT and have a pixel scale of $0.5$\arcsec\ at Solar Orbiter (corresponding to $198$ Mm on the Sun), a cadence of $180$ s, and a rotation angle (compared to IRIS) of approximately $-1.26^\circ$. Once again, jitter within the time-series was removed using the $tr\_get\_disp.pro$ IDL routine, before a $701$$\times701$ pixel sub-FOV corresponding to the IRIS FOV was extracted from each of the sixteen larger $2048$$\times2048$ pixel maps. A plot of the sub-FOV extracted in the previous steps (saturated at $\pm500$ G, where white indicates positive polarity and black indicates negative polarity) as measured at approximately $10$:$24$:$13$ UT can be seen in the left-hand panel of Fig.~\ref{Overview}.

\subsubsection{SDO}
\label{SDO}

Finally, we also studied data sampled by both the Atmospheric Imaging Assembly (SDO/AIA; \citealt{Lemen12}) and the Helioseismic and Magnetic Imager (SDO/HMI; \citealt{Scherrer12}) instruments on-board SDO. These instruments constantly sample the entire solar disk, as observed from Earth, with (post-reduction) pixel scales of $0.6$\arcsec\ ($435$ km). We note that we used the $hmi\_prep.pro$ routine to convert SDO/HMI data onto the equivalent SDO/AIA orientation and pixel scaling. Here, we studied images sampled by various UV ($1600$ \AA\ and $1700$ \AA) and EUV ($304$ \AA, $171$ \AA, and $211$ \AA) SDO/AIA filters with $24$ s and $12$ s cadences, respectively, and  maps of the line-of-sight magnetic field returned by SDO/HMI with cadences of $45$ s. Sub-fields of $401$$\times401$ pixels were extracted from each of these datasets at each time-step, with the returned FOV approximately matching the IRIS FOV. Importantly, these data allow for both a better investigation of the thermal evolution of the studied features as well as a closer alignment between different instruments.

\begin{figure*}
\includegraphics[width=0.99\textwidth]{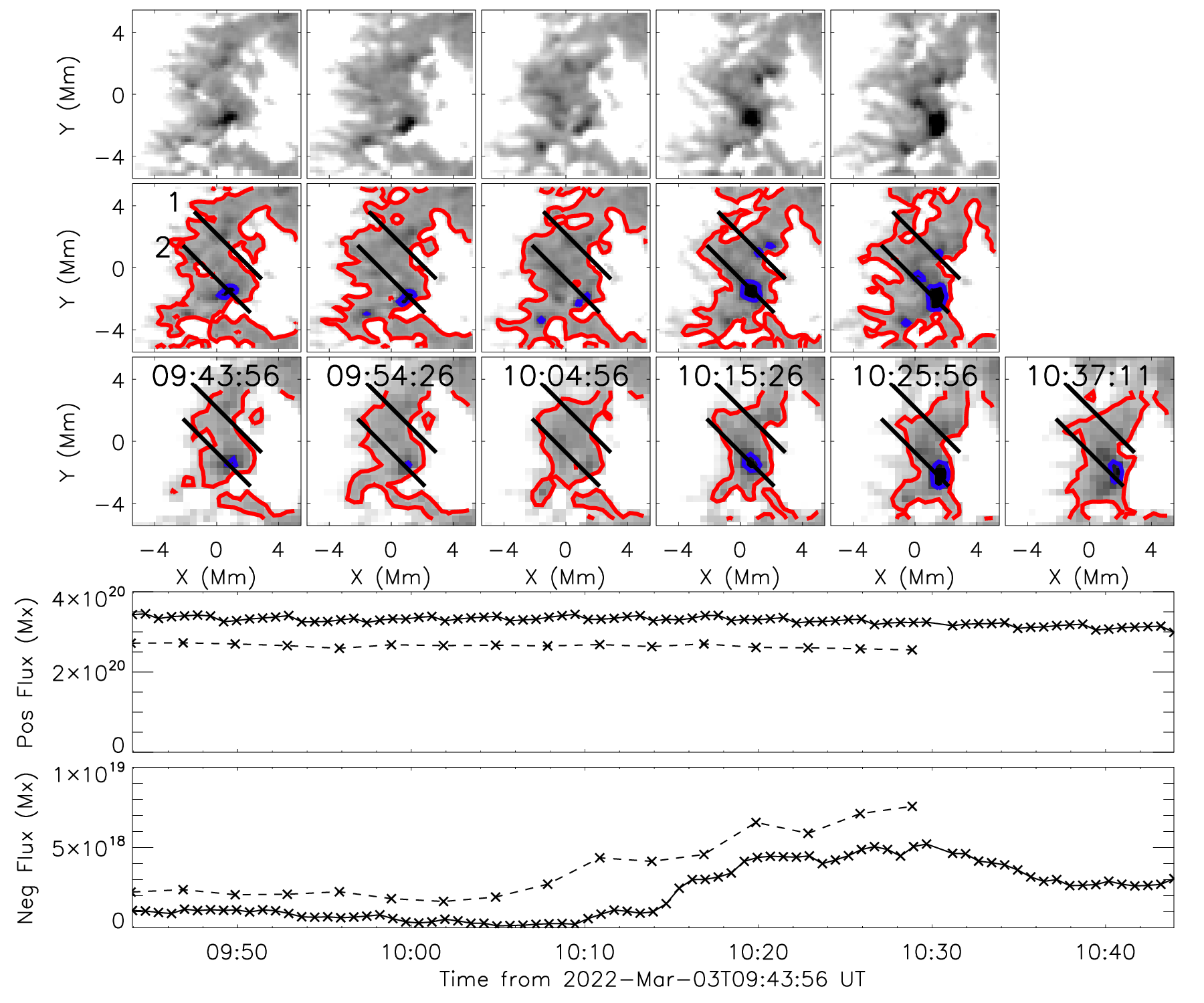}
\caption{Evolution of the line-of-sight magnetic field within the blue box over-laid on Fig.~\ref{Overview}. The top row plots the FOV as sampled by SO/PHI-HRT, saturated at $\pm200$ G, for five time-steps (the approximate times of which are over-laid on the third row). The second row plots the same as the top row but is annotated with contours outlining regions of $\pm100$ G (red for positive and blue for negative polarities) and two lines (labelled as `1' and `2') that are used to construct the time-distance diagrams plotted in Fig.~\ref{Mag_ts}. The third row plots the FOV as measured by the SDO/HMI instrument, with the annotations being the same as in the second row. The fourth and fifth rows plot the evolution of the total positive polarity flux (fourth row) and negative polarity flux (fifth row) within the FOV through time as measured by SDO/HMI (solid line) and SO/PHI-HRT (dashed line). The crosses indicate individual measurements. Note the differing scales on the y-axes of these plots. We note that the step-like evolution of the SDO/HMI flux, particularly evident in the positive polarity flux plot, is an artefact introduced by the single-pixel derotation applied to those data through time. A movie more clearly displaying the evolution of this region (combined with Fig.~\ref{LA_evol}) is included with the online version of this article.}
\label{Mag_evol}
\end{figure*}

\subsection{Alignment}
\label{Alignment}

\begin{figure*}
\includegraphics[width=0.99\textwidth]{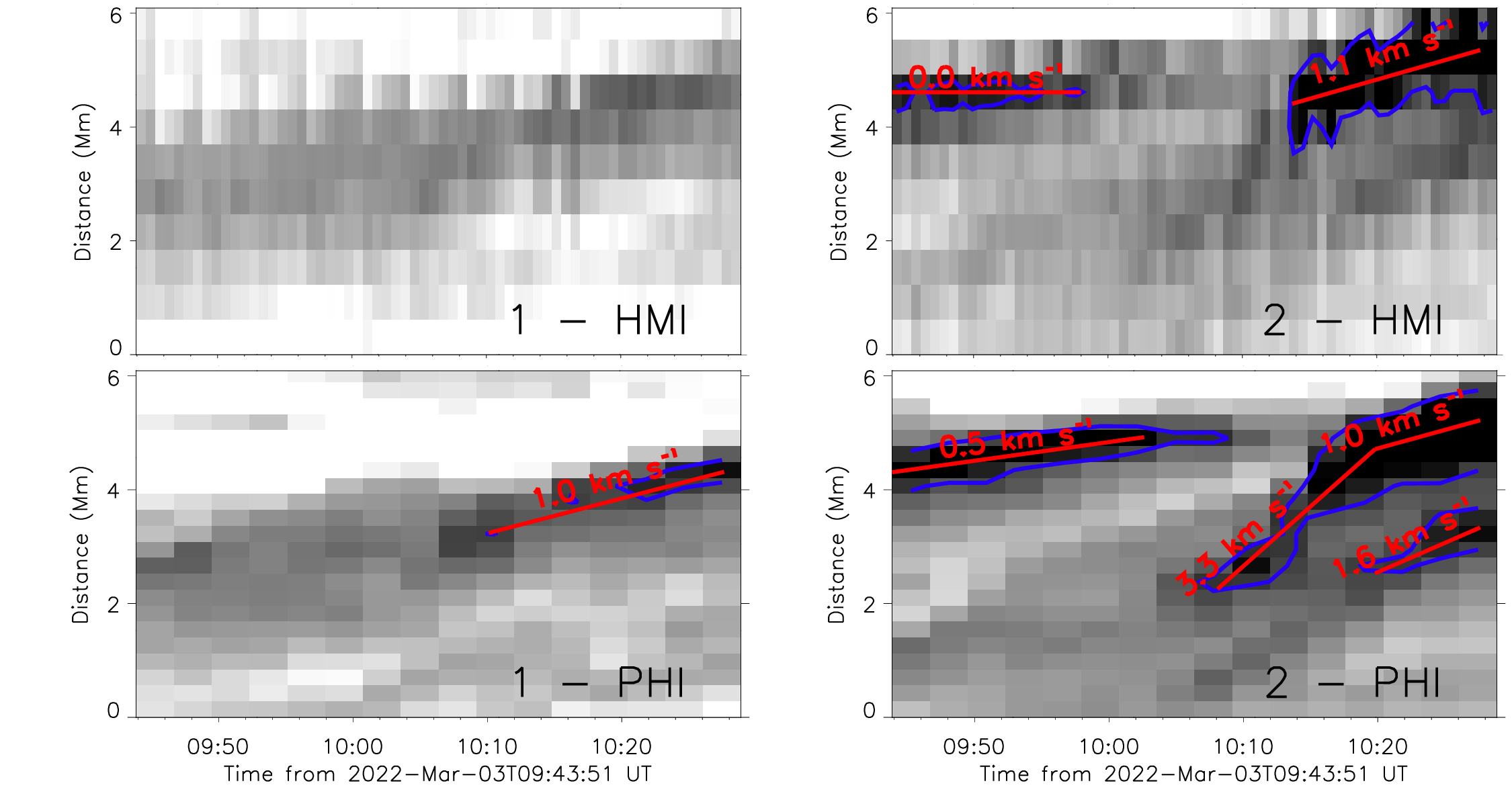}
\caption{Time-distance diagrams constructed along lines `1' (left-hand panels) and `2' (right-hand panels) over-laid on Fig.~\ref{Mag_evol} for both SDO/HMI (top panels) and SO/PHI-HRT (bottom panels). The blue contours outline regions where the negative polarity MMFs are above an absolute value of $100$ G, whilst the annotated red lines indicate the approximate speeds of these negative polarity MMFs.}
\label{Mag_ts}
\end{figure*}

Given the relative differences in position between Solar Orbiter and the other satellites used here, alignment between data sampled by each individual instrument is non-trivial. Specifically, the IRIS SJI, SO/PHI-HRT, and \HRIEUV\ data all sampled different heights of the solar atmosphere so any direct co-alignment between features would not return accurate results. In order to analyse all of the datasets in a consistent manner, therefore, we used SDO as a central resource against which Solar Orbiter and IRIS could both be aligned. For each instrument, we began the alignment process by calculating the position of each pixel within the FOV in Carrington coordinates. This was done using the $euvi\_heliographic.pro$ IDL routine and by considering the header information to be a ground-truth. We then identified the pixels in each dataset that were closest, again assuming no errors in the header information, in space to the centre of the IRIS FOV for the respective frame closest to the first IRIS SJI image. After this, sub-FOVs for each dataset around that value were constructed for each dataset. Next, we created contour maps of each dataset and blinked each IRIS SJI and Solar Orbiter dataset against the most relevant SDO dataset. Any off-sets evident at this stage were removed by shifting the dataset with the highest spatial resolution by an integer number of pixels in the $x$- and $y$- direction. This step was repeated for each figure plotted in this article. As the datasets from each instrument were already aligned to themselves through time, we only conducted cross-instrument alignments at the beginning of the time-series, with the results applied throughout. Specifically, the first frame of the \HRIEUV\ image sequence was aligned to the image sampled the closest temporally by the SDO/AIA $171$ \AA\ filter, the first SO/PHI-HRT line-of-sight magnetic field map was aligned to the temporally closest SDO/HMI line-of-sight magnetic field map, and the first IRIS SJI $1400$ \AA\ image was aligned to the temporally closest SDO/AIA $1600$ image. Overall, we consider that the alignment after these steps was closer than $0.6$\arcsec\ (i.e. better than the pixel scale of the SDO instruments post-reduction).

\section{Results}
\label{Results}

\subsection{Magnetic field evolution}  

We begin our research by analysing the evolution of the line-of-sight magnetic field within the blue box over-laid on Fig.~\ref{Overview}. Given that SO/PHI-HRT data have a higher spatial resolution whilst SDO/HMI data have a higher temporal resolution, we compare both here to gain a more complete understanding of the dynamics of this region. In the top three rows of Fig.~\ref{Mag_evol}, we plot measurements of the line-of-sight magnetic field within this box at numerous time-steps during this time-series. The top row plots a sub-FOV of the SO/PHI-HRT data at five intervals separated by approximately $12$ minutes, with the magnetic field values saturated at $\pm200$ G. This sub-FOV is clearly dominated by positive polarity field, with the edge of the sunspot encroaching on the left-hand side of the sub-FOV whilst a large ($\sim10$ Mm diameter) region of positive polarity field exists on the right-hand side. Between these two positive polarity regions, several small ($0.5$-$2$ Mm diameter) negative polarity moving magnetic features (MMFs) are evident, moving from top-left to bottom-right (i.e. away from the sunspot) through time (the movement of these features is most clearly seen in the associated movie, as well as in Fig.~\ref{Mag_ts}). The second row plots the same as the top row, but with contours over-laid at a level of $\pm100$ G to highlight the negative polarity MMFs and to serve as a reference for later figures. Additionally, two lines (annoted as `1' and `2') display the paths of several small negative polarity MMFs through time. The third row plots the same FOV, equivalently annotated, as sampled by the SDO/HMI instrument, with an additional column corresponding to a time when SO/PHI-HRT had stopped observing. The most immediately evident difference is that the smaller negative polarity MMFs are not observable in SDO/HMI (see, for example, the third column where no negative polarity MMF with absolute field strength above $100$ G are measured).   

\begin{figure*}
\includegraphics[width=0.99\textwidth]{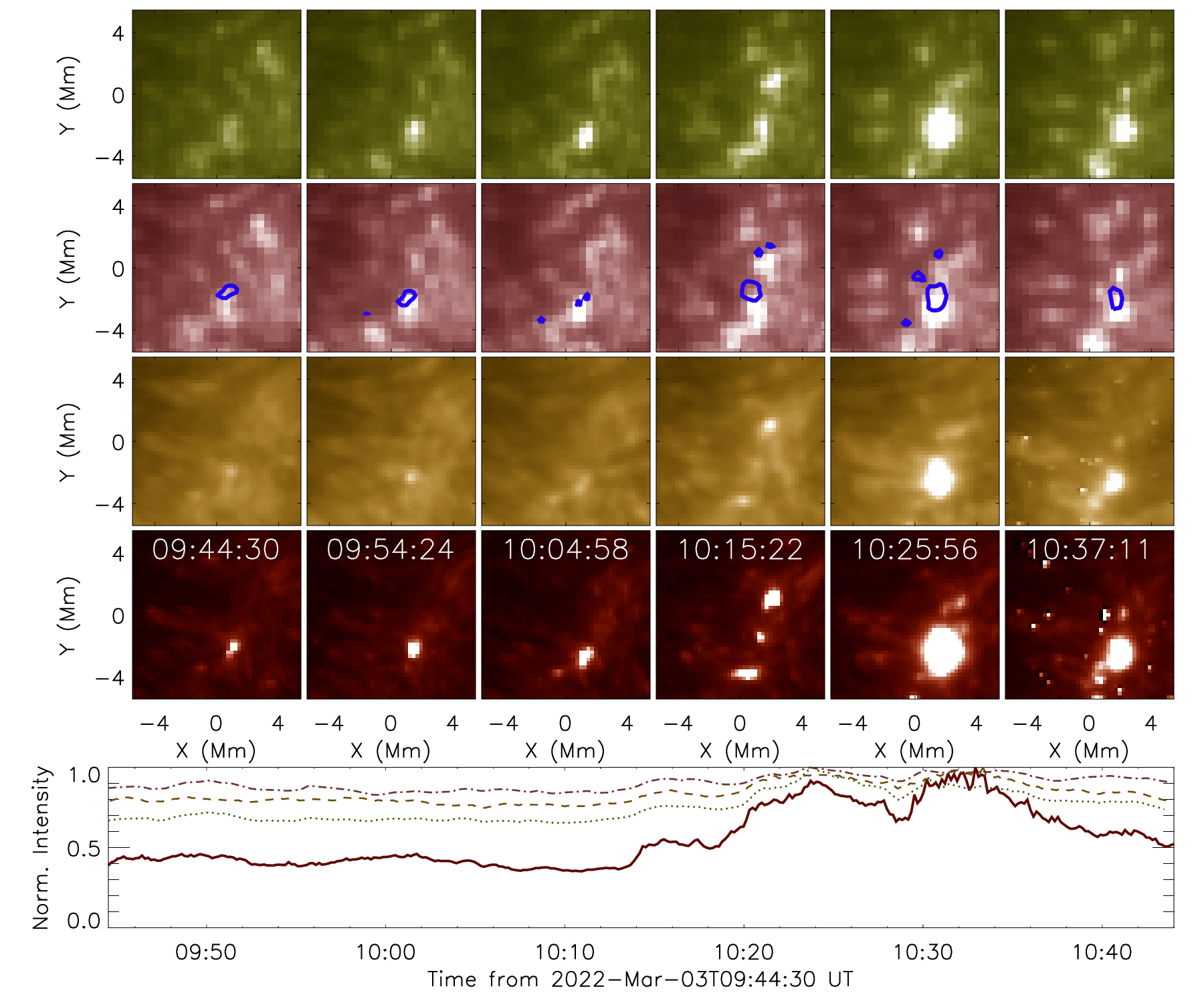}
\caption{Temporal evolution of the UV Bursts identified here in four UV channels, namely the SDO/AIA $1600$ \AA\ channel (top row), SDO/AIA $1700$ \AA\ channel (second row), IRIS SJI \ion{Mg}{II} $2796$ \AA\ channel, and IRIS SJI \ion{Si}{IV} $1400$ \AA\ channel. The blue contours on the second row outline the co-spatial line-of-sight magnetic field at a value of $-100$ G as sampled by SO/PHI-HRT (first five columns) and SDO/HMI (sixth column). Multiple UV Bursts are immediately evident in the IRIS SJI \ion{Si}{IV} $1400$ \AA\ data through time. The bottom row plots the evolution of the intensity within this FOV through time for the IRIS SJI \ion{Si}{IV} $1400$ \AA\ channel (solid line), the IRIS SJI \ion{Mg}{II} $2796$ \AA\ channel (dashed line), the SDO/AIA $1600$ \AA\ channel (dotted line), and the SDO/AIA $1700$ \AA\ channel (dot-dashed line). A movie more clearly displaying the evolution of this region (combined with Fig.~\ref{Mag_evol}) is included with the online version of this article.}
\label{LA_evol}
\end{figure*}

In the bottom two rows of Fig.~\ref{Mag_evol}, we plot the total magnetic flux measured within this FOV through time for both the positive flux (fourth row) and negative flux (fifth panel). The solid lines are measured from SDO/HMI data, whilst the dashed lines are measured from SO/PHI-HRT data. The positive polarity flux is two order of magnitude larger than the negative polarity flux and remains relatively stable through time (for both SO/PHI-HRT and SDO/HMI). A slight reduction in the total positive polarity flux (at a rate of $1.4$$\times10^{16}$ Mx s$^{-1}$) is measured, however, it is impossible to ascertain whether this is due to changes in the morphology of the large positive polarity features (e.g. the sunspot moving in and out of the FOV) or due to dynamic processes (e.g. emergence or cancellation) occurring within the FOV. The evolution of the total negative polarity flux is, though, easier to tie to the evolution of the visible features in the FOV. Initially, around $2.5$$\times10^{18}$ Mx of negative polarity flux is measured within the FOV, with the majority of this being contained within a single MMF close to the centre of the FOV. Around $10$:$10$ UT, though, several new negative polarity MMFs are measured between the sunspot and the large region of positive polarity flux on the right of the FOV. This increase in flux occurs at a rate of around $4.2$$\times10^{15}$ Mx s$^{-1}$ and predominantly takes place along the paths indicated by the labelled lines on the second row. We note that there are slight differences in the total flux values obtained by both SO/PHI-HRT and SDO/HMI (around $20$ \%\ for the positive polarity). These are likely due to the combined effects of differences in the areas of the FOVs measured over (due to the differing pixel scales), some small differences in the instruments themselves (e.g. sampling different wavelength positions across the line), and the different lines-of-sight of the satellites.

\begin{figure*}
\includegraphics[width=0.99\textwidth]{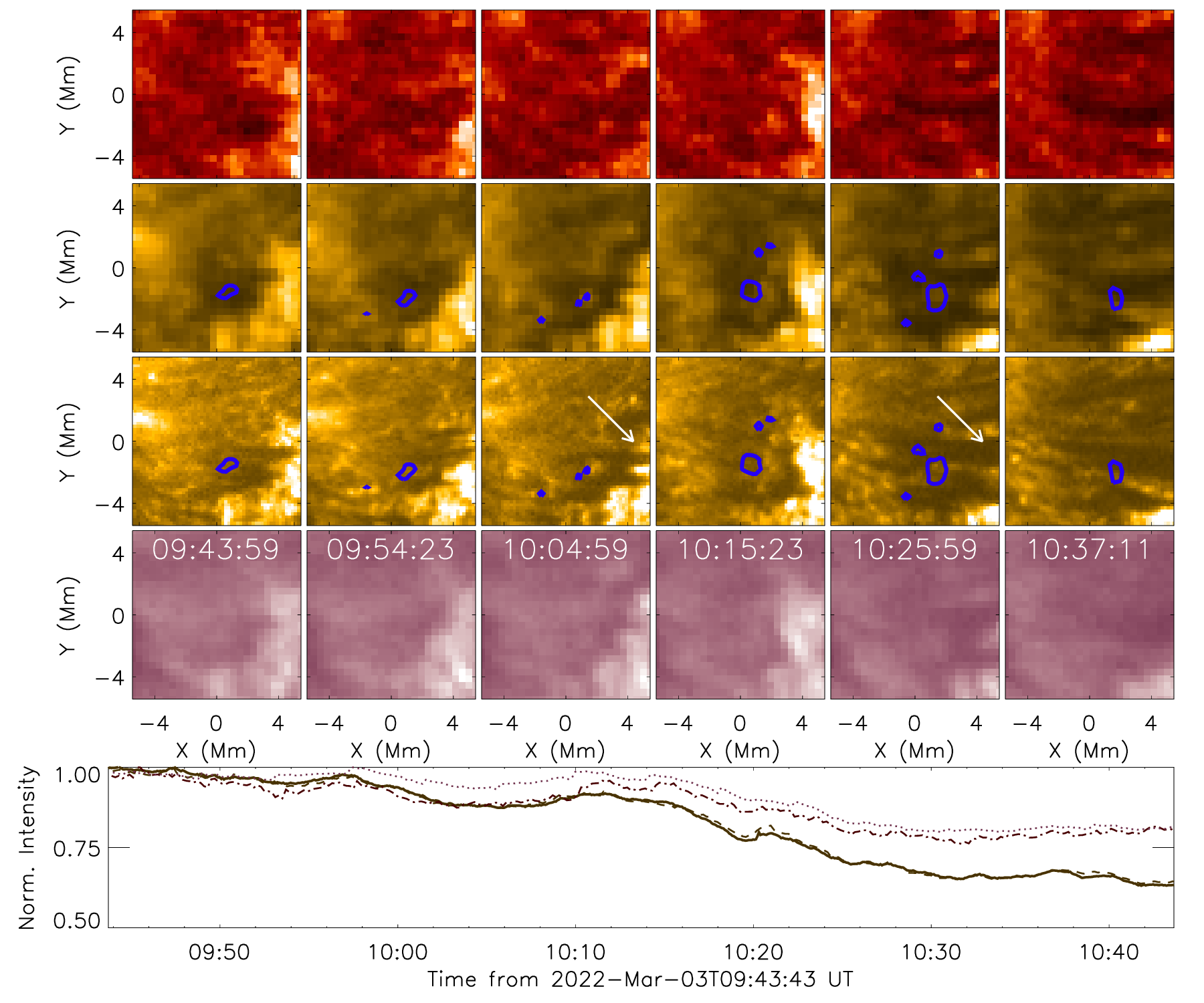}
\caption{Same of for Fig.~\ref{LA_evol} but for the SDO/AIA $304$ \AA\ channel (top row - dot-dashed line), SDO/AIA $171$ \AA\ channel (second row - dashed line), \HRIEUV\ data (third row - solid line), and SDO/AIA $211$ \AA\ channel (fourth row - dotted line). No localised brightenings are detected co-spatial to the UV Bursts identified previously, however, a reduction in the normalised mean intensity was seemingly associated with the ejection of surges from the reconnection site. The two arrows on the \HRIEUV\ row indicate the first short surge extension (third column) and the second larger surge (fifth column), discussed in Sect.~\ref{Surges}, that occurred co-spatial to the central UV Burst within this FOV.}
\label{UA_evol}
\end{figure*}

In order to further investigate the motions of the evolving negative polarity MMFs, we constructed time-distance diagrams along the annotated lines plotted on the second and third rows of Fig.~\ref{Mag_evol} for the time-period sampled by SO/PHI-HRT. In the left-hand column of Fig.~\ref{Mag_ts}, we plot time-distance diagrams constructed along line `1' for both SDO/HMI (top panel) and SO/PHI-HRT (bottom panel). The dark blue contours outline the regions of negative polarity field with absolute values above $|100|$ G that comprise the MMFs, whilst the red lines are used to approximate the speed of these MMFs through time. For line `1', no clear negative polarity MMF is identified in the SDO/HMI data, however, a small MMF is detected in SO/PHI-HRT data, moving away from the sunspot between $10$:$10$ UT and $10$:$28$ UT at a speed of around $1.0$ km s$^{-1}$. For line `2' (right-hand panels), two negative polarity MMFs are evident in the SDO/HMI data with the first appearing to remain stationary over the course of the first $15$ minutes sampled and the second moving at around $1.1$ km s$^{-1}$ radially outwards from the sunspot between $10$:$10$ UT and $10$:$28$ UT. The SO/PHI-HRT time-distance diagram reveals additional dynamics along line `2'. Notably, the stationary negative polarity MMF in the SDO/HMI data is now observed to move with a speed of $0.5$ km s$^{-1}$, whilst the second negative polarity MMF is observed to potentially have a very high-speed component between $10$:$08$ UT and $10$:$18$ UT of $3.3$ km s$^{-1}$ (much higher than the speeds of typical MMFs; \citealt{Hagenaar05,Qin19}), prior to a deceleration to a speed of around $1.0$ km s$^{-1}$ that is sustained until $10$:$28$ UT. We use the word `potentially' here when referring to $3.3$ km s$^{-1}$ motions as the relatively low cadence of the SO/PHI-HRT data means we are not able to fully rule out other possibilities (such as multiple concurrent regions of emergence occurring along the slit) that could explain this high-speed component. However, the movies included with the online version of this article do give the impression of motions, at least to these authors. Additionally, a third negative polarity MMF is detected in the SO/PHI-HRT data  (with a speed of $1.6$ km s$^{-1}$) occurring behind the very high-speed component of the second MMF. Interestingly, therefore, despite having a lower cadence, the higher spatial resolution of the SO/PHI-HRT data allows us to more clearly track the motions of small-scale MMFs through time (a similar result has been shown previously using ground-based data by \citealt{Reid16}).

\begin{figure*}
\includegraphics[width=0.99\textwidth]{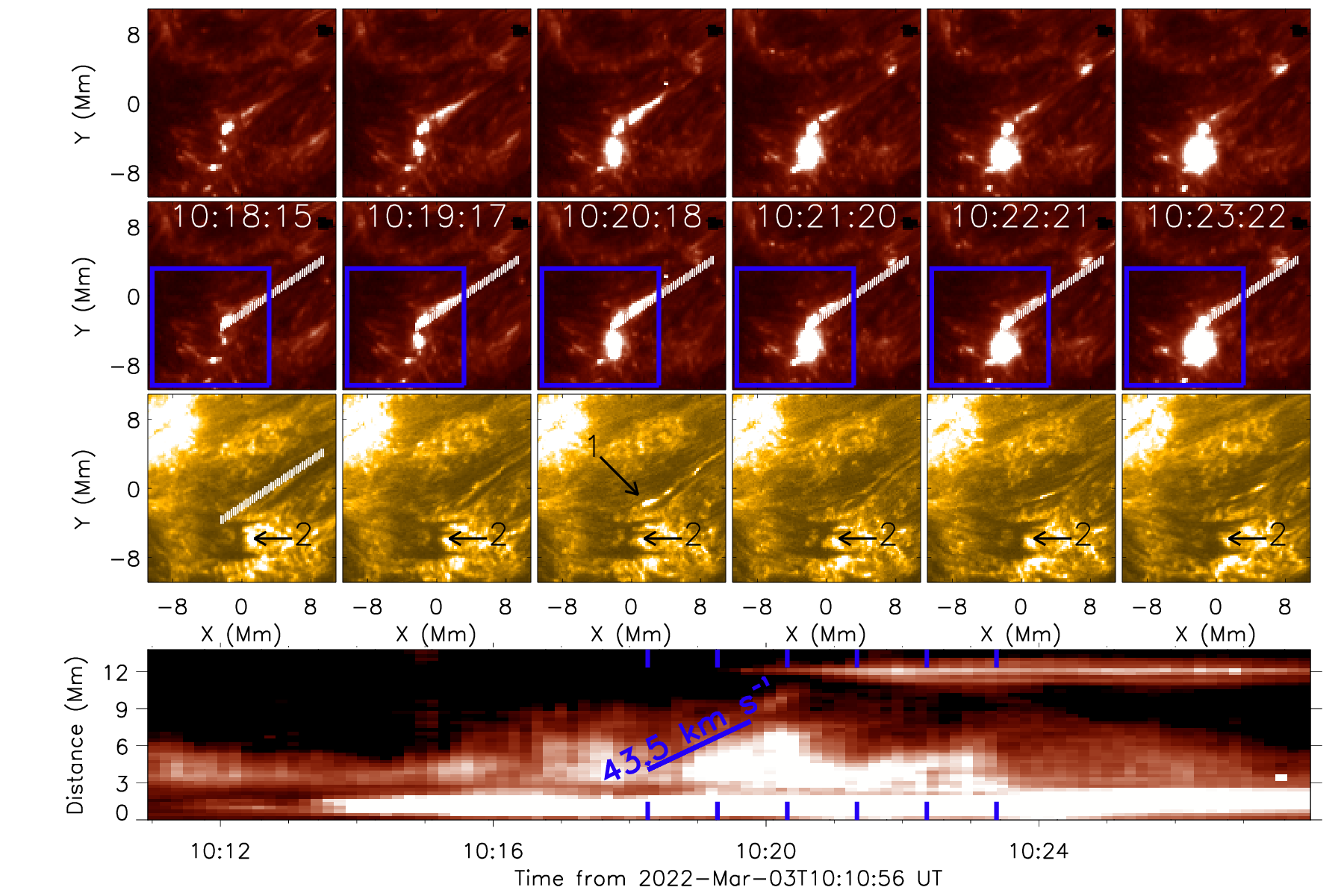}
\caption{Evolution of the surge associated with the northern most UV Burst identified in Fig.~\ref{LA_evol} over the course of around $5$ minutes. The top panel plots the IRIS SJI $1400$ \AA\ intensity, the second row plots the same but with a blue box indicating the FOV plotted in Fig.~\ref{LA_evol} and white lines indicating the pixels averaged over to create a time-distance diagram over-laid, whilst the third row plots the co-spatial and co-temporal \HRIEUV\ intensity. The white lines over-laid on the first column of the third row are exactly co-spatial to the white lines over-laid on the first column of the second row, whilst the black arrows indicate the locations of the two surges related to the northern most (`1') and central (`2') UV Bursts studied here. The bottom panel plots a time-distance diagram constructed using IRIS SJI \ion{Si}{IV} $1400$ \AA\ data at the pixels identified by the white lines on the second row. The six blue markers at the top and bottom indicate the time-steps plotted above. The extension of the surge is highlighted by the diagonal blue line that has a gradient of $43.5$ km s$^{-1}$. A movie more clearly displaying the evolution of this region (combined with Fig.~\ref{Surge2_evol}) is included with the online version of this article.}
\label{Surge1_evol}
\end{figure*}

\subsection{UV Bursts}

To begin our investigation into the response of the local plasma to the dynamics observed in the line-of-sight magnetic field, we analysed the evolution of the intensities of four spectroscopic channels sampling the lower solar atmosphere (typical formation temperatures $\le8$$\times10^4$ K) within the FOV out-lined by the blue box over-laid on Fig.~\ref{Overview}. The top four rows of Fig.~\ref{LA_evol} plot the SDO/AIA $1600$ \AA\ (top row) and $1700$ \AA\ (second row) channels, as well as the IRIS SJI \ion{Mg}{II} $2796$ \AA\ (third row) and \ion{Si}{IV} $1400$ \AA\ channels at six time-steps. The blue contours over-laid on the SDO/AIA $1700$ \AA\ images outline the negative polarity field islands at a value of $-100$ G identified in Fig.~\ref{Mag_evol}, with the first five columns being from SO/PHI-HRT data and the final column being from SDO/HMI data. Initially, a small UV Burst (diameter of around $1$-$2$ Mm) is apparent in the IRIS SJI $1400$ \AA\ channel co-spatial to the polarity inversion line between a small negative polarity MMF and the larger positive polarity plage region at the right-hand side of the FOV. This UV Burst remains relatively stable in the IRIS SJI $1400$ \AA\ data over the first $20$ minutes of this time-series before fading from view at around $10$:$07$ UT. This UV Burst is also present in the other channels studied, being particularly bright between $09$:$59$ UT and $10$:$03$ UT, however, it is not as clearly visible as in the IRIS SJI $1400$ \AA\ channel (see the movie associated with Figs.~\ref{Mag_evol} and ~\ref{LA_evol}). Overall, this evolution is consistent with the hypothesis that magnetic reconnection between the negative polarity magnetic island identified to move at a speed of $0.5$ km s$^{-1}$, in the bottom right panel of Fig.~\ref{Mag_ts}, and the more stable positive polarity region on the right of the FOV is driving this UV Burst.

Co-temporal to the initial detection and motions of the new negative polarity MMFs between the sunspot and the positive polarity plage (between $10$:$08$ UT and $10$:$20$ UT), three UV Bursts develop in the \ion{Si}{IV} $1400$ \AA\ channel (see fourth column of Fig.~\ref{LA_evol}). The northern most of these occurs along the axis of line `1' from Fig.~\ref{Mag_evol}, becoming visible at around $10$:$13$ UT, peaking in size (at around $2.2$ Mm) and intensity at around $10$:$23$ UT, before finally fading from view around $10$:$40$ UT. The central UV Burst detectable in the fourth column occurs close to line `2' from Fig.~\ref{Mag_evol}, where the seemingly high-speed moat flows are detected. This UV Burst expands rapidly from a diameter of around $0.7$ Mm at $10$:$14$ UT to a diameter of more than $3.6$ Mm at $10$:$23$ UT, much larger than typical UV Bursts (see \citealt{Young18}), and remains around that size until the end of the time-series. The southern most UV Burst is more intermittent, occurring three times (with diameters of around $0.7$ Mm) between $10$:$14$ UT and the end of the time-series. Again, each of these UV Bursts is detectable in the other channels studied, but is most clear in the \ion{Si}{IV} $1400$ \AA\ data. The bottom panel of Fig.~\ref{LA_evol} plots the normalised mean intensity across this FOV through time for each of the four channels studied here. The solid line plots the \ion{Si}{IV} $1400$ \AA\ intensity which shows much higher contrast than the other channels, as would be expected given the increased visibility of the UV Bursts in those images. We note that spiked pixels during the transition of IRIS through the South Atlantic Anomoly (SAA; see the sixth column, for example) were removed before the calculation of the mean intensity. The strong UV Bursts occurring during the initial detection and movement of the negative polarity MMFs are identifiable as large peaks between $10$:$13$ UT and the end of the time-series. As with the UV Burst detected earlier in the time-series, the evidence suggests that magnetic reconnection between the newly measured negative polarity MMFs and the positive polarity plage region on the right of the FOV is driving all of these events.

\begin{figure*}
\includegraphics[width=0.99\textwidth]{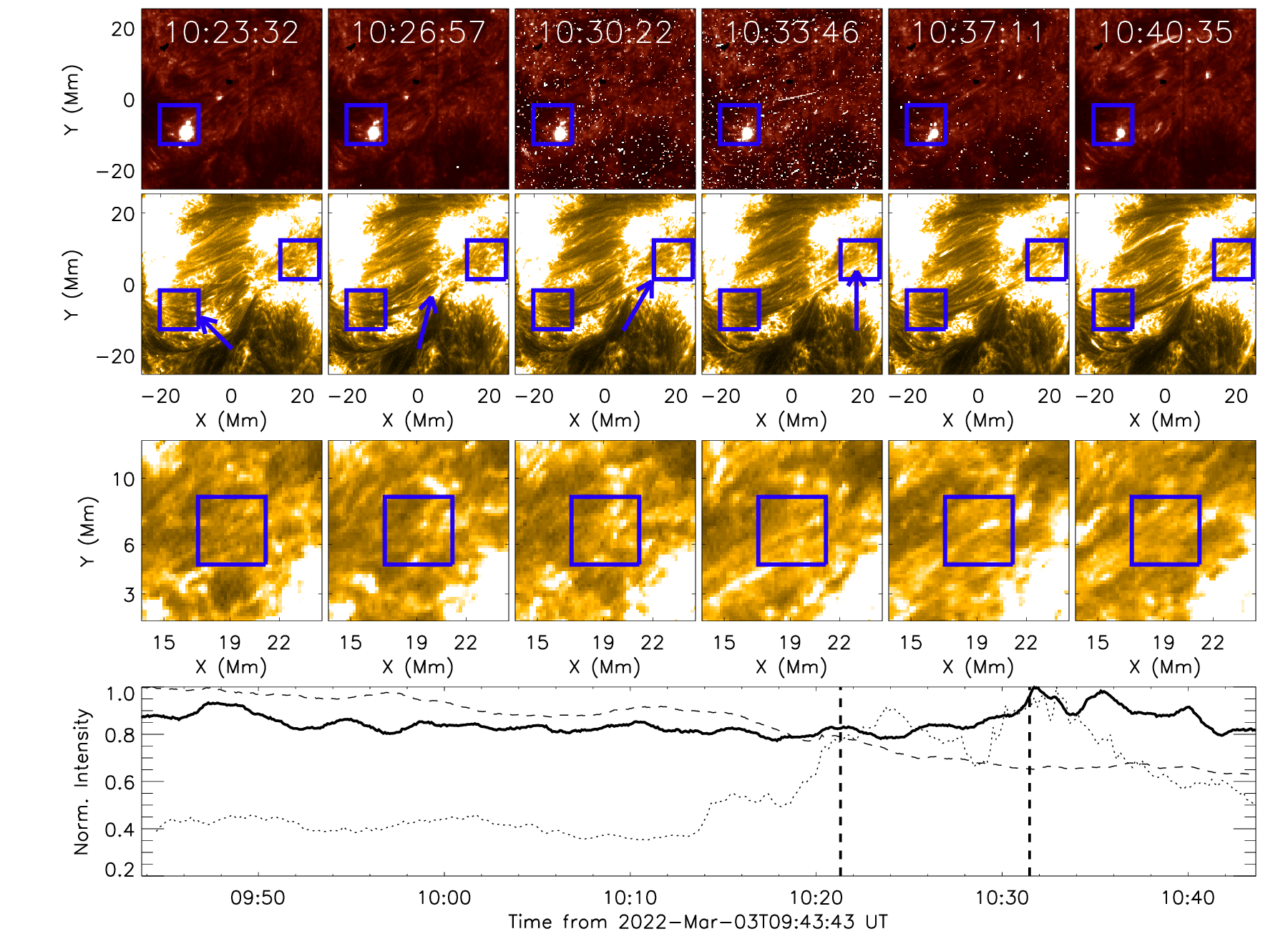}
\caption{Evolution of the large surge associated with the central UV Burst identified in Fig.~\ref{LA_evol}. The top row plots the IRIS SJI \ion{Si}{IV} $1400$ \AA\ intensity at six time-steps, with the over-laid blue box indicating the FOV plotted in Fig.~\ref{LA_evol}. The second row plots the co-spatial and co-temporal FOV as sampled by the \HRIEUV\ telescope. The right-most blue box indicates the location of the other foot-point of the surge, whilst the over-laid blue arrows indicate the approximate location of the tip of the surge through time. The third row plots co-temporal \HRIEUV\ images sampled within the right-hand box over-laid on the second row. The over-laid blue box outlines the region where the intensity is averaged over to construct a lightcurve of this foot-point. The bottom panel plots this light curve (solid line), as well as the IRIS SJI \ion{Si}{IV} $1400$ \AA\ (dotted) and \HRIEUV\ (dashed) intensities measured around the UV Bursts at the left foot-point. The first vertical dashed line indicates the apparent start time of the surge at the left foot-point whilst the second vertical dashed line indicates the time at which the surge appears to reach the right-hand foot-point. A movie more clearly displaying the evolution of this region (combined with Fig.~\ref{Surge1_evol}) is included with the online version of this article.}
\label{Surge2_evol}
\end{figure*}

Typically, UV Bursts do not display any signature in coronal channels (see, for example, \citealt{Young18}), however, with the increased spatial and temporal resolution offered by the \HRIEUV\ data it was of interest to confirm this here, particularly given the large size of the central UV Burst. In Fig.~\ref{UA_evol}, we plot the equivalent as Fig.~\ref{LA_evol} but for the SDO/AIA $304$ \AA\ (top row), SDO/AIA $171$ \AA\ (second row), \HRIEUV\ (third row), and SDO/AIA $211$ \AA\ data. Once again, the blue over-laid contours outline the negative polarity MMFs as identified from SO/PHI-HRT (first five columns) and SDO/HMI (sixth column). As expected, no compact brightenings are detectable in the coronal channels during this time. Plotting light-curves of these channels (bottom panel) did reveal some interesting behaviour though. Specifically, the mean intensity in all four studied channels dropped throughout this time-series with the largest gradients being detectable after $10$:$10$ UT. This reduction in mean intensity is particularly evident in the \HRIEUV\ (solid line) and SDO/AIA $171$ (dashed line) time-series, with a drop to around $60$ \%\ of the original intensity being measured. Studying the \HRIEUV\ images, it became apparent that the reduction in mean intensity was caused by the ejection of dark surge material, seemingly from the reconnection site, that lead to increased absorption at the right of the FOV. Expanding the FOV studied, it also became evident that a bright surge was emitted from the northern-most UV Burst earlier in the time-series. We study both of these surges in detail in the following sub-section.

\begin{figure*}
\includegraphics[width=0.99\textwidth]{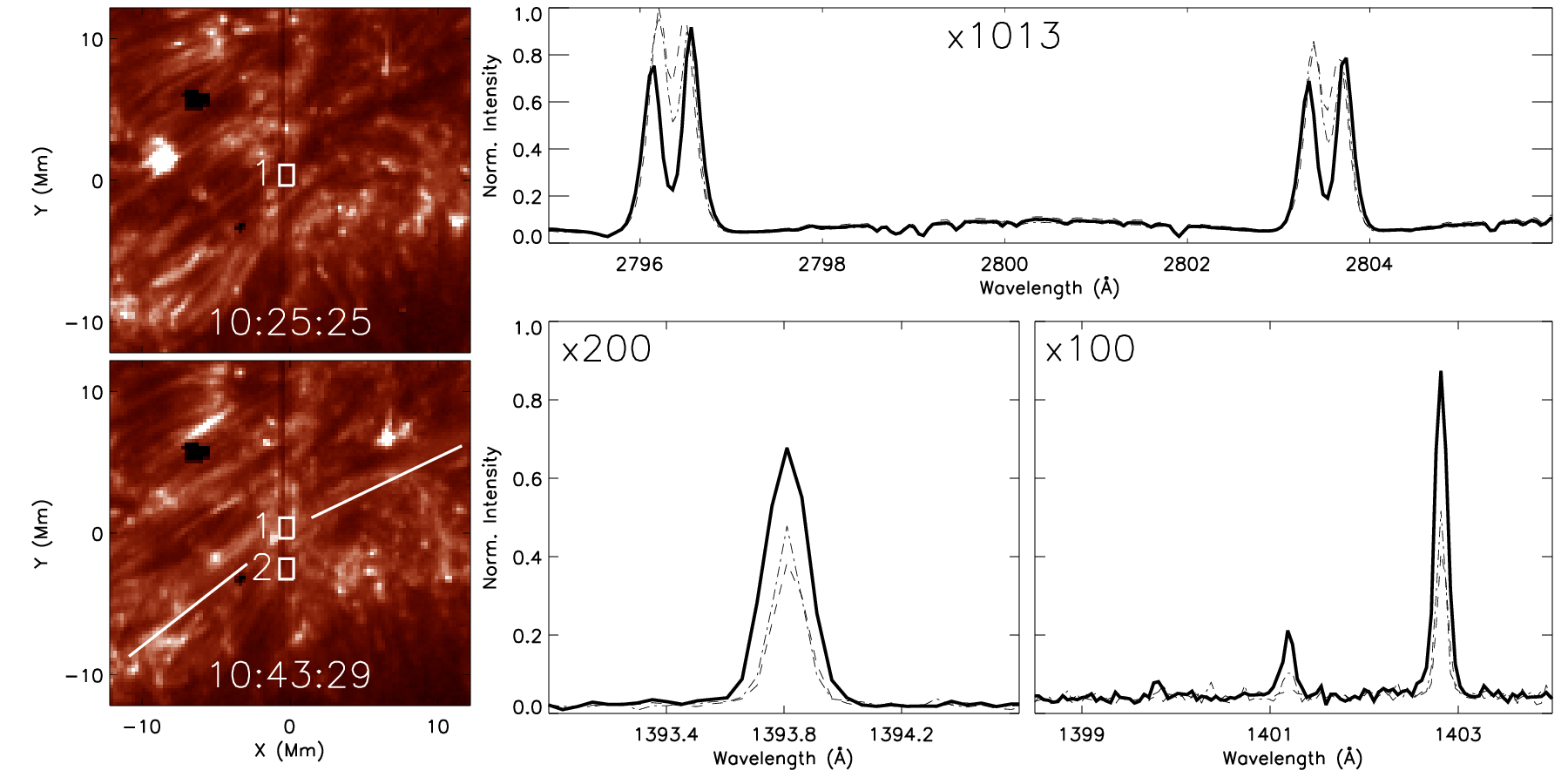}
\caption{Differences between the IRIS spectra sampled before and after the South Atlantic Anomaly (SAA), corresponding to before and after the initial ejection of the surge. The left-hand panels plot the local FOV as sampled by the IRIS SJI $1400$ \AA\ filter before (top panel) and after the SAA (bottom panel). The white boxes labelled `1' indicate the location of the surge on the spectrograph slit, whilst the box labelled `2' is taken as a quiet background area. The two white lines indicate the approximate path of the surge across the FOV. The right hand panels plot spectra sampled in the \ion{Mg}{II} windows (top panel), \ion{Si}{IV} $1394$ \AA\ window (bottom left panel), and \ion{Si}{IV} $1403$ \AA\ window (bottom right panel). The solid lines plot the spectra measured from box `1' during the surge, whilst the dashed and dot-dashed lines plot the spectra measured from box `1' before the surge and from box `2', respectively. The specific multiplication factors for each plot are over-laid for reference.}
\label{IRIS_Spec}
\end{figure*}

\subsection{Surges}
\label{Surges}

We begin our analysis of the surges by studying the feature apparently linked to the northern most UV Burst detected in Fig.~\ref{LA_evol}. In Fig.~\ref{Surge1_evol}, we plot the evolution of this event over the course of around $5$ minutes. The surge can be detected as a bright ejection from the UV Burst in the IRIS SJI \ion{Si}{IV} $1400$ \AA\ data (top row), pointing towards the top right corner. This surge evolves quickly extending to a length of around $5$ Mm between the first and third columns before seeming to retract again. The second row plots the same as the top row but with annotations over-laid, including a blue box that outlines the FOV plotted in Fig.~\ref{LA_evol} and white lines that indicate the pixels averaged over to construct the time-distance diagram plotted in the bottom panel. The third row plots the co-spatial and co-temporal \HRIEUV\ intensity. The same white bars are plotted on the first column to help guide the eye. Although neither a bright surge nor a dark surge is seen at this location in this channel, several short (lifetimes of $20$-$30$ s) duration compact brightenings are observed along the length of the surge at different times. A clear example of this is indicated by the arrow labelled with a `1' over-laid on the third column of Fig.~\ref{Surge1_evol}. We, therefore, suggest that the main body of this surge is heated to transition region temperatures (around $8$$\times10^4$ K), with the sub-structuring in the \HRIEUV\ data potentially indicating that additional transient heating to coronal temperatures (around $10^6$ K) also takes place at localised regions. Studying the time-distance diagram constructed from the IRIS SJI $1400$ \AA\ intensity along the length of the slit (bottom panel), we can clearly identify the bright ejection propagating away from the northern most UV Burst (situated  between $0$-$2.9$ Mm) between $10$:$18$ UT and $10$:$24$ UT. The plane-of-sky speed of this surge is approximately $43.5$ km s$^{-1}$, as highlighted by the annotated blue line. The co-formation of this surge with the northern most UV Burst indicates that this event could also be driven by magnetic reconnection between the negative polarity MMF, that moves along line `1' from Fig.~\ref{Mag_evol}, and the positive polarity plage region to the right of the FOV. 

The event associated with the central UV Burst identified in Fig.~\ref{LA_evol}, and hence the seemingly high-speed $3.3$ km s$^{-1}$ MMF, is very different to the surge associated with the northern UV Burst. Initially (between around $10$:$10$ UT and $10$:$20$ UT), a short ejection of dark surge material, with a peak length of around $3.6$ Mm can be identified propagating away from the location of the central UV Burst. This short ejection moves over a region of bright plage causing increased absorption and hence accounts for the first clear reduction in intensity apparent in the bottom panel of Fig.~\ref{UA_evol}. The retraction of this event around $10$:$20$ UT is then responsible for the slight increase in intensity measured around that time. The subtle dynamics of this initial short surge ejection are highlighted by the black arrows labelled with a `2' on Fig.~\ref{Surge1_evol}. Following this, a longer dark surge is then ejected from the same spatial location at around $10$:$22$ UT. In the top two rows of Fig.~\ref{Surge2_evol}, we plot the evolution of this event over the course of around $17$ minutes in the IRIS SJI $1400$ \AA\ (top row) and \HRIEUV\ (second row) data over a much larger FOV ($51$$\times51$ Mm$^2$). The large and relatively stable central UV Burst is immediately evident within the blue box over-laid on the top row (that outlines the FOV plotted in Fig.~\ref{LA_evol}); however, only a very faint ejection (that cannot be seen in these plotted images but does become apparent when investigating the movies of the event included with the online version of this article) linked to the surge is visible in data from this channel. From this same position in the \HRIEUV\ data, though, a clear dark surge emerges and propagates towards the blue box over-laid on the right-hand side of the FOV. The blue arrows indicate the tip of the surge as it progresses from left to right across the FOV. These signatures in spectroscopic data suggest that this surge is largely comprised of cool (potentially chromospheric) material that is leading to increased absorption in the \HRIEUV\ data. 

The third row of Fig.~\ref{Surge2_evol} plots a zoom-in of the FOV outlined by the blue box on the right of the second row. No evidence of the surge is present in the first three columns, but after that a diffuse bright structure can be identified connecting the central blue box to the bottom left of the FOV. This structure remains until the end of the time-series, suggesting that this region is the other foot-point of a closed loop within which the surge propagates. Plotting the intensity of the central blue box over-laid on the third row (bottom panel - solid line), we find a slight increase in intensity at around $10$:$32$ UT indicating the arrival of the surge material (agreeing with dynamics observed in \HRIEUV\ images). The dotted and dashed lines re-plot the intensity as measured around the UV Bursts from the IRIS SJI \ion{Si}{IV} $1400$ \AA\ and \HRIEUV\ datasets, respectively. The first vertical dashed line indicates the start time of the surge at the left foot-point whilst the second vertical dashed line indicates the first evidence of the arrival of surge material at the right foot-point around $610$ s later. Given the distance between the two blue boxes is around $36.8$ Mm, we can calculate a plane-of-sky speed of the surge to be approximately $60.3$ km s$^{-1}$. However, if instead we assume that the surge material follows the path of a semi-circle, modelling a potential magnetic loop connecting the two foot-points in the photosphere, then we can calculate the speed of the surge material to be around $94.7$ km s$^{-1}$ (consistent with speeds of surges reported previously; see, for example, \citealt{Nobrega17,Nobrega21}). 

Interestingly, the larger surge associated with the central UV Burst crosses across the IRIS spectrograph slit. Unfortunately, the majority of the dynamics of this event occur during the passage of IRIS through the South Atlantic Anomaly (SAA) meaning we are limited to comparisons of spectra from before and after this time. In Fig.~\ref{IRIS_Spec}, the left-hand panels plot before (top) and after (bottom) the SAA for a FOV around the surge as sampled by the IRIS SJI $1400$ \AA\ channel. The box labelled `1' sits at the location of the surge, whilst the box labelled `2' is a local quiet region that we class as the background. The two diagonal white lines plot the approximate path of the surge across the FOV. The right-hand panels plot the average spectra measured within the boxes on the left-hand panels. The background spectra and the spectra measured at the surge location before the SAA are plotted by the dot-dashed and dashed lines, respectively. These spectra are remarkably similar in each of the three wavelength windows plotted. The spectra measured from within the surge after the SAA are plotted by the solid lines. We note four differences between the surge spectra and the background spectra here. Firstly, the \ion{Mg}{II} spectra (top panel) within the surge after the SAA displays the opposite asymmetry to the background and before the SAA spectra. Secondly, the \ion{Si}{IV} $1394$ \AA\ and $1403$ \AA\ lines both increase in intensity with only marginal line broadening and Doppler velocities. Thirdly, the line ratio between the \ion{Si}{IV} spectra ($1394$/$1403$) decreases below $2$, indicating that the plasma becomes more optically thick when the surge is present. Finally, both the \ion{O}{IV} $1399$ \AA\ and \ion{O}{IV} $1401$ \AA\ lines are detectable when the surge is present with line ratios that imply an electron density of between $10^{11}$-$10^{12.5}$ cm$^{-3}$ for this event (see \citealt{Young18b}). We note that the \ion{O}{IV} $1399$ \AA\ line is weak (even after averaging over multiple pixels) and, therefore, we do not try to provide a more accurate estimate of the electron density. However, this value does match qualitatively with those reported previously for surges (see \citealt{Nobrega21}).

\section{Discussion}
\label{Discussion}

\subsection{Identification of small-scale magnetic features}

The primary motivation for including both SO/PHI-HRT and SDO/HMI line-of-sight magnetic field maps in our research was to allow a comparison of the obtained results. Although it is known that these two instruments return comparable line-of-sight magnetograms (see the in-depth analysis conducted by \citealt{Sinjan23}), small-scale bursts and brightenings in the solar atmosphere typically have lifetimes below $10$ minutes (see, for example, \citealt{Nelson15,Young18,Berghmans21,Nelson23} amongst many others) and, as such, it was possible that the lower cadence of the SO/PHI-HRT data (despite having better than double the spatial resolution, at $0.547$ AU where these data were sampled, of SDO/HMI) may lead to important temporal dynamics being missed. Our analysis showed the opposite, however, with the dynamics of the line-of-sight magnetic field being much more clearly observed in the SO/PHI-HRT measurements (see Figs.~\ref{Mag_evol}-\ref{Mag_ts} and the movies included with the online version of this article). Indeed, the SO/PHI-HRT data allowed us to potentially identify a very high-speed (compared to typical MMFs as measured by \citealt{Hagenaar05,Qin19}) of $3.3$ km s$^{-1}$ , for one of the negative polarity MMFs as it moved away from the sunspot. This motion was not easily identifiable in SDO/HMI data (see Fig.~\ref{Mag_ts}), however, it may have become evident if we considered lower thresholds for the detection of such events. The fact that SDO/HMI and SO/PHI-HRT both measured the lower speed (around $1$ km s$^{-1}$) of this feature at later times gives us confidence that the initial higher speed measurement from SO/PHI-HRT data is correct. Using lower magnetic field strength thresholds for identifying the MMFs may reveal similar behaviour in SDO/HMI data, however, it is clear from the plots that SO/PHI-HRT is better capable of identifying such motions.

One notable difference between the results obtained from the two different instruments lay in the calculated magnetic flux values (see bottom two panels of Fig.~\ref{Mag_evol}). For both the positive and the negative flux, there were differences in the calculated values with no specific trend being noted. Indeed, SDO/HMI returned larger positive and total unsigned flux values, whilst SO/PHI-HRT returned larger negative flux values within the small FOV plotted in Fig.~\ref{Mag_evol}. It is likely that these differences are caused by the differing FOVs measured over (due to the differing spatial resolutions of the instruments), subtle differences between the instruments (for example, in terms of wavelengths sampled), the different lines-of-sight (LOS) of the two satellites, and the different codes employed to extract the LOS field component from the measured Stokes parameters. Over the entire AR studied these effects appear to be small, with the total unsigned magnetic fluxes measured by the two instruments being $1.20$$\times10^{22}$ Mx for SO/PHI-HRT and $1.27$$\times10^{22}$ Mx for SDO/HMI. Additionally, if we consider the rate of change of the magnetic flux (as is common when analysing small-scale events; see, for example, \citealt{Reid16,Nelson16}) within the small FOV plotted in Fig.~\ref{Mag_evol} then this appears to be well correlated between instruments for both the positive and negative fluxes. Performing a cross correlation (using $c\_correlate.pro$) of the negative flux values returned by both SO/PHI-HRT and SDO/HMI (when both instruments were observing), for example, returns a coefficient of 0.92. Our research suggests, therefore, that the rate of change of the magnetic flux may be a good metric to use if one wishes to combine SO/PHI-HRT and SDO/HMI line-of-sight magnetic field maps (e.g., to extend observations of a specific FOV beyond a SO/PHI-HRT observing window).

\subsection{Formation of associated UV Bursts and surges}

Understanding the connectivity between small-scale bursts identified in the UV with those identified across other temperature regimes in the solar atmosphere is a complex task. IRIS Bursts (\citealt{Peter14}), for example, are identified in \ion{Si}{IV} $1394$ \AA\ spectra which have formation temperatures of around $8$$\times10^4$ K; however, these events have been reported to sometimes form co-spatially with both EBs (\citealt{Vissers15,Chen19}) and QSEBs (\citealt{Nelson17}), which are identified in the wings of the H$\alpha$ spectral line and are thought to have formation temperatures below $10^4$ K. On the other hand, EEs (\citealt{Brueckner83}) also have formation temperatures close to $10^5$ K but have been observed to form co-spatially with EUV brightenings (\citealt{Nelson23}) in spectroscopic \HRIEUV\ data, which are predominantly sensitive to plasma with temperatures closer to $10^6$ K. Attempts to create unified models that account for these varied signatures across multiple thermal windows are on-going but have had some success, with the effects of magnetic topology (e.g., \citealt{Hansteen19}) and plasma beta (e.g., \citealt{Peter19}) seemingly being important variables that define whether such features are observed individually or in combination. 

The research conducted here highlights an important issue to consider when creating a unifying model of small-scale bursts. Specifically, various manifestations of small-scale bursts have been identified at the foot-points of surges in the solar atmosphere (\citealt{Roy73,Madjarska09,Watanabe11,Reid15,Nelson19,Nobrega21}), meaning any unifying model must also account for this occasional linkage. Not only this, but surges themselves also have varying thermal responses, as well as spatial and temporal scales, that must be taken into account. In the data studied here, we find both a short, hot surge (transition region temperatures; Fig.~\ref{Surge1_evol}) and a long, cool surge (presumably chromospheric temperatures; Fig.~\ref{Surge2_evol}) being driven from the same $7$$\times7$ Mm$^2$ region within a few minutes of each other. It is certainly plausible that magnetic topology and the local plasma beta could also play a role in defining whether a surge is formed co-spatial to a small-scale burst and, thereafter, its spectroscopic visibility. Previous research, for example, has suggested that the interaction between emerging flux and pre-existing ambient flux could be important in driving surges (\citealt{Yokoyama95,Guglielmino10}). Another important factor could be the local photospheric dynamics such as the strength of the horizontal velocities (as discussed by \citealt{Kitai10}). The region studied here included both high horizontal velocities (potentially measured as high as $3.3$ km s$^{-1}$ from SO/PHI-HRT data) as well as interactions between emerging (negative polarity) low-lying flux islands and a (positive polarity) larger-scale (around $36$ Mm diameter) ambient loop system, potentially creating perfect conditions for both small-scale UV burst and surge formation.

\section{Conclusions}
\label{Conclusions}

In this article, we have studied the formation of UV Bursts and associated surges that formed in the moat flow around a sunspot in AR $12957$ (Fig.~\ref{Overview}). We have used a multi-instrument approach (analysing spectroscopic imaging data, line-of-sight magnetic field maps, and spectral data) to comprehensively constrain the properties of these events. We found that:
\begin{itemize}
\item{The SO/PHI-HRT line-of-sight magnetic field maps revealed higher plane-of-sky speeds for the small-scale negative polarity MMFs around the sunspot, when compared against SDO/HMI (see Fig.~\ref{Mag_evol}). These speeds (potentially up to $3.3$ km s$^{-1}$ radially away from the sunspot; Fig.~\ref{Mag_ts}) are higher than those observed for typical MMFs (\citealt{Hagenaar05,Qin19}) and last long enough to be measured despite the relatively low cadence ($180$ s) of SO/PHI-HRT line-of-sight magnetic field maps. A follow up study analysing the speeds of a larger statistical sample of MMFs around several sunspots in SO/PHI-HRT data, particularly observed close to perihelion when the spatial resolution is highest, would be useful. An increased temporal cadence during SO/PHI-HRT observations would also help to shed light on these events.}
\item{The UV Bursts, identified most easily in IRIS SJI $1400$ \AA\ data, formed at polarity inversion lines located between the negative polarity MMFs and a larger-scale positive polarity plage region around $7.2$ Mm away from the sunspot. These UV Bursts were, generally, larger (diameters of up to $3.6$ Mm) and longer-lived (lifetimes of over $20$ minutes) than typical such events (see \citealt{Young18}) in spectroscopic data sampling from the photosphere to the transition region (Fig.~\ref{LA_evol}). However, they had no signature in imaging data sampling the corona (Fig.~\ref{UA_evol}). This spectroscopic coverage confirms these events as UV Bursts rather than hotter events such as, for example, micro-flares. We were unable to further classify these events (e.g., into IRIS Bursts or EEs) due to the lack of spectral information at these locations.}
\item{Multiple surges were also observed to form around the polarity inversion lines identified in SO/PHI-HRT data, with the two clearest examples being very different in nature. The first surge studied was relatively short (length around $5.1$ Mm), had a plane-of-sky speed of $43.5$ km s$^{-1}$, and was seemingly heated to transition region temperatures (Fig.~\ref{Surge1_evol}). The second surge studied was relatively long (connecting foot-points around $36.8$ Mm apart), had an apparent speed of around $94.6$ km s$^{-1}$ (assuming the surge traced a potential semi-circular loop connecting the two foot-points), and was visible as a dark feature in \HRIEUV\ data (Fig.~\ref{Surge2_evol}). No downward motions were observed in this surge suggesting that this event consisted of chromospheric material that filled a closed loop, before presumably slowly draining at a later time. Interestingly, this surge crossed the IRIS spectrograph slit allowing comparisons between the spectra before and after the formation of the event. Our results indicated that the surge was optically thick and had a local electron density of between $10^{11}$-$10^{12.5}$ cm$^{-3}$ in the region of the solar atmosphere sensitive to the \ion{O}{IV} spectral lines sampled by IRIS.}
\end{itemize}
Combining these conclusions obtained through studying different regions of the solar atmosphere allows us to propose a scenario that explains this apparent linkage between MMFs, UV bursts, and surges. Here, we find fast MMFs impacting onto an apparent region of locally open magnetic field (connecting 10s of Mm away), which creates UV bursts in the lower solar atmosphere and drives surges into the upper solar atmosphere, filling a large loop structure. We propose, therefore, that magnetic reconnection between a fast-moving low-lying magnetic field topology and a more stable coronal loop-like topology may create the conditions required for UV burst and surge linkages. This scenario is qualitatively similar to that discussed in simulations by \citet{Yokoyama95}. Overall, these results clearly show the benefits of multi-instrument analyses when it comes to small-scale events, and offer new insights into the relationship between UV Bursts and surges.

\begin{acknowledgements}
CJN is thankful to ESA for support as an ESA Research Fellow. This research was supported by the International Space Science Institute (ISSI) in Bern, through ISSI International Team project \#23-586 (`Novel Insights Into Bursts, Bombs, and Brightenings in the Solar Atmosphere from Solar Orbiter'). This project has received funding from the European Research Council (ERC) under the European Union's Horizon 2020 research and innovation programme (grant agreement No. 101097844 — project WINSUN). Solar Orbiter is a mission of international cooperation between ESA and NASA, operated by ESA. We are grateful to the ESA SOC and MOC teams for their support. The German contribution to SO/PHI is funded by the BMWi through DLR and by MPG central funds. The Spanish contribution is funded by AEI/MCIN/10.13039/501100011033/ and European Union “NextGenerationEU”/PRTR” (RTI2018-096886-C5,  PID2021-125325OB-C5,  PCI2022-135009-2, PCI2022-135029-2) and ERDF “A way of making Europe”; “Center of Excellence Severo Ochoa” awards to IAA-CSIC (SEV-2017-0709, CEX2021-001131-S); and a Ramón y Cajal fellowship awarded to DOS. The French contribution is funded by CNES. The EUI instrument was built by CSL, IAS, MPS, MSSL/UCL, PMOD/WRC, ROB, LCF/IO with funding from the Belgian Federal Science Policy Office (BELSPO/PRODEX PEA 4000134088, 4000112292, 4000136424, and 4000134474); the Centre National d’Etudes Spatiales (CNES); the UK Space Agency (UKSA); the Bundesministerium f\"ur Wirtschaft und Energie (BMWi) through the Deutsches Zentrum f\"ur Luft und Raumfahrt (DLR); and the Swiss Space Office (SSO). IRIS is a NASA small explorer mission developed and operated by LMSAL with mission operations executed at NASA Ames Research Center and major contributions to downlink communications funded by ESA and the Norwegian Space Centre. SDO/AIA and SDO/HMI data provided courtesy of NASA/SDO and the AIA and HMI science teams. This research has made use of NASA’s Astrophysics Data System Bibliographic Services.
\end{acknowledgements}

\bibliographystyle{aa}
\bibliography{UV_Burst_Surges}

\end{document}